\documentclass{article}
\usepackage{sao2,psfig}
\newcommand {\hs}{\hspace*{1.5mm}}   
\issue{1999, 48, 5-16}
\begin{document}

\title{
     5 radio sources of the Zenith survey at RATAN-600:
     VLA
\thanks{The National Radio Astronomy Observatory is operated under license by
Associated Universities, Inc. under cooperative agreement with
the National Science Foundation.
}
maps, radio spectra and optical identification
}

\author{Yu.N. Parijskij\inst{a}
\and W.M. Goss \inst{b}
\and O.V. Verkhodanov \inst{a}
\and A.I. Kopylov \inst{a}
\and\\ N.S. Soboleva \inst{c}
\and A.V. Temirova \inst{c}}

\institute{\saoname
\and National Radio Astronomical observatory,
      P.O. Box 0, Socorro, New Mexico 87801, USA
\and St.Petersburg branch of the Special Astrophysical Observatory,
     Pulkovo, St.Petersburg, Russia
}
\date{March 19, 1999}{September 1, 1999}
\maketitle

\begin{abstract}
VLA
maps obtained at 1.4 GHz
with a resolution of 2\farcs5$\times$2\arcsec~for three sources and
6\farcs5$\times$2\farcs3 for two sources
detected  in the RATAN--600 Zenith survey of 1988
have been analyzed. All five objects have an extended structure.
Continuous radio spectra of these objects prepared by using
the CATS database and RATAN-600 observations are given.
All five objects have linear steep spectra ($\alpha < -0.65$).
Using APM database and DSS we have found three candidates for identification.
Two radio sources have been observed with the 6 m telescope.
The nature of the studied objects, one of which is classified as QSO (BSO)
and two others as galaxies, is discussed.
One of the radio sources, $RZ5$, being a merging group of one large and
several small galaxies, may have appeared in this process.
\end{abstract}

\section{Introduction}

In 1988 a survey was carried out at the RATAN--600 in the
right ascension range between $8^h$ and $14^h$
at declination 47\degr7\arcmin~of 1\arcmin~ width with
the use of the entire ring surface of the radio telescope
(Mingaliev et al., 1991). A catalog of 70 objects ($RZ$ sources)
has been compiled at the wavelength of 8.0~cm. All the detected
sources belong to a population of radio sources with flux densities
from 14 to 70 mJy.
General statistical investigation, estimation of spectral indices
and luminosity calculation (in the assumption of $z\sim$1) of
$RZ$ sources have been reported by Verkhodanov (1994).
Several sources of this survey were studied using new
RATAN--600 observational data obtained by Verkhodanov and Verkhodanova
(1997--1999) in 1995 at four wavelengths.

To reveal morphology and select candidates for
study with optical telescopes, one has to investigate
the radio structure of the objects. By using morphological properties
and selecting FRII objects (Fanaroff and Riley, 1974), one could
extend the lists of the objects being candidates for distant radio
galaxies studied in the programme ``Big Trio'' (Parijskij et al., 1996).
Detection and investigation of this type objects play an important role
in the understanding of the origin and evolution of galaxies
at early epochs of the Universe.
By studying
faint radio sources that are close doubles (down to 2\arcsec)
we can also select gravitationally lensed candidates which belong
to the most interesting objects in the present day astrophysics
(Fletcher, 1998).
These objects can be found among radio sources extended but
unresolved with a beam of 2\farcs5.

\begin{table*}
\caption{VLA data for RZ sources}
\begin{center}
\begin{tabular}{|lccllrrccr|}
\hline
Name&$\alpha+\delta(J2000)$&$\alpha+\delta(B1950)$&$\sigma_{\alpha}$&$\sigma_{\delta}$&S&$\sigma_S$&Major&Minor&PA\\
      &      &     &s&\arcsec&mJy & mJy&\arcsec&\arcsec&\degr \\
\hline
RZ5N  & 082346.18+465200.3&082014.22+470142.0&.02 & .1 &  34 &  4 & 5.37 & 4.28 & 134.3 \\
RZ5S  & 082347.37+465148.6&082015.42+470130.4&.01 & .1 & 110 & 15 & 8.85 & 4.46 & 158.6 \\
RZ5int& 082347.13+465150.4&082015.18+470132.2&.04 & .6 & 180 & 15 & 23   &      & 141   \\
RZ9   & 084141.23+465234.5&083812.66+470318.2&.01 & .1 &  70 &  7 & 3.97 & 1.64 & 120.2 \\
RZ14  & 084818.21+465153.1&084451.01+470258.7&.02 & .1 &  53 &  4 & 5.58 & 0.74 &  24.6 \\
RZ55  & 131217.54+465106.0&131005.66+470659.8&.01 & .1 & 150 & 20 & 9.24 & 1.33 & 164.7 \\
RZ70  & 135751.31+465130.5&135552.48+470604.7&.01 & .1 & 180 & 40 & 5.50 & 0.66 & 146.5 \\
\hline
\end{tabular}
\end{center}
\end{table*}

\begin{figure*}
\centerline{\psfig{figure=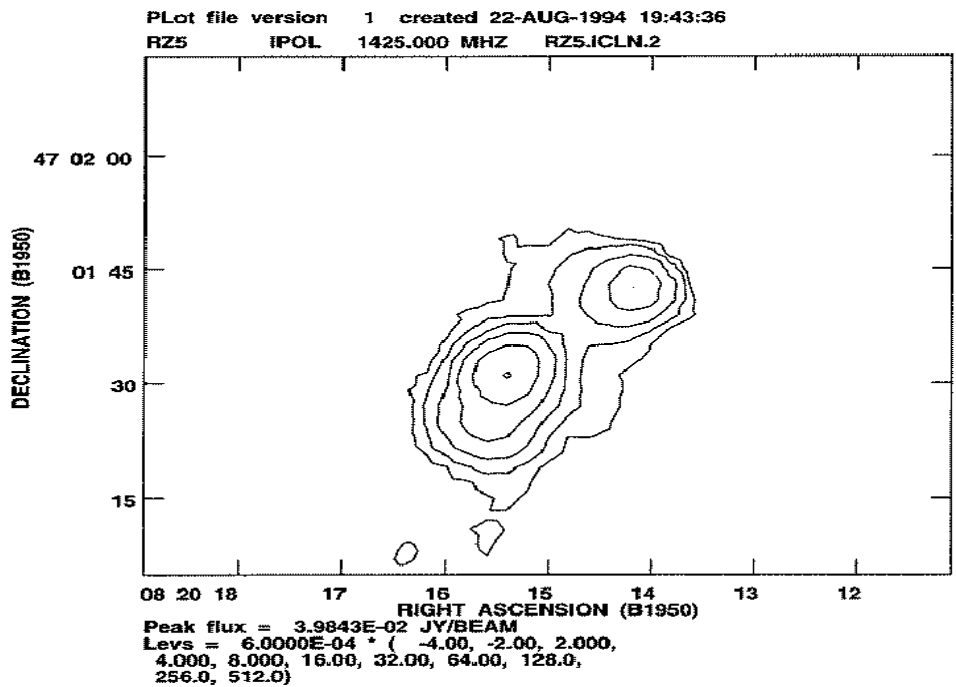,width=14cm,height=10cm,angle=-90}}
\caption{Radio source RZ5 (J~082347+465150).}
\end{figure*}

We studied the structure of 5 sources of the Zenith survey
for the purpose of revealing morphology
from the images obtained
with VLA
 in intermediate configurations (BnA) in 1994.
Besides, we have obtained continuous radio spectra of the objects  using
the new radio sky surveys NVSS (New VLA Sky Survey), FIRST (Faint Images
of the Radio Sky at Twenty-cm) and WENSS (The Westerbork Northern Sky
Survey), and the data of
other surveys. We have made identification of the studied sources
on the optical images of the Digital Sky Survey (DSS) and with the
Automized Plate Measuring machine (APM) database via Internet.
Two $RZ$ objects have been observed at the 6 m telescope.
Classification of all five objects has been carried out after the optical
and radio investigation.

\section{VLA maps}

Maps of three Zenith survey radio sources, {\it RZ5} (Fig.1),
{\it RZ9} (Fig.2) and {\it RZ14} (Fig.3), with a resolution
of 2{\farcs}5$\times$2\arcsec~at 1425 MHz
and of two sources, {\it RZ55} (Fig.4) and {\it RZ70} (Fig.5),
with a resolution of 6\farcs5$\times$2\farcs3~at
1455 MHz were obtained with VLA in 1994.
These objects were selected from the general $RZ$ list as the brightest
ones with steep spectra
($\alpha < -0.6$, S$\sim\nu^\alpha$) and proposed to the VLA observations in
the unified list of the ``Big Trio'' objects.
Isophotes shown on the maps of these sources (Figs. 1-5) are drawn by levels
proportional to a factor of 2 starting
with 1.2, 1.0, 0.6, 1.0, 1.2 mJy, respectively.
Positive isophotes are shown by
the solid lines, and negative by the dotted lines.
Table 1 contains the data for these objects.
In the columns are given the  object name,
coordinates at the epoch of 2000.0 and coordinate errors, flux densities
and their errors in mJy, deconvolved major and minor axes of
the radio sources in arcsec,
positional angle in degrees.
Coordinates of the integrated source ($RZ5int$) have been
taken from the NVSS data.
The major and minor axes and the positional angle  for all sources have
been borrowed from the FIRST catalog.

\begin{figure*}
\centerline{\psfig{figure=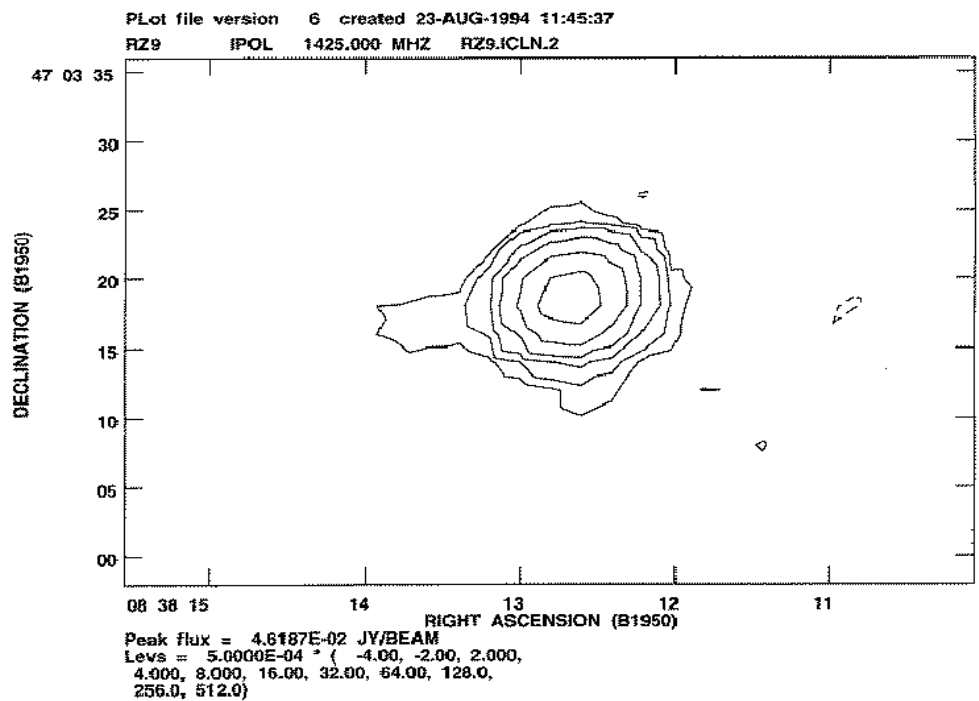,width=14cm,height=10cm,angle=-90}}
\caption{Radio source RZ9 (J~084141+465234).}
\end{figure*}

\begin{figure*}
\centerline{\psfig{figure=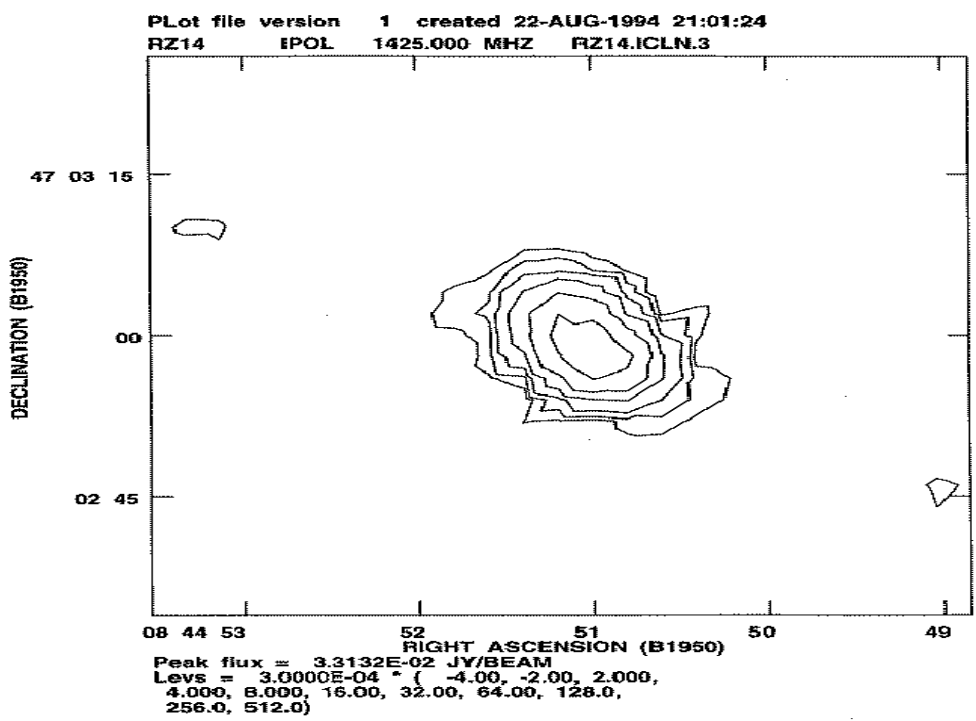,width=14cm,height=10cm}}
\caption{Radio source RZ14 (J~084818+465153).}
\end{figure*}

\begin{figure*}
\centerline{\psfig{figure=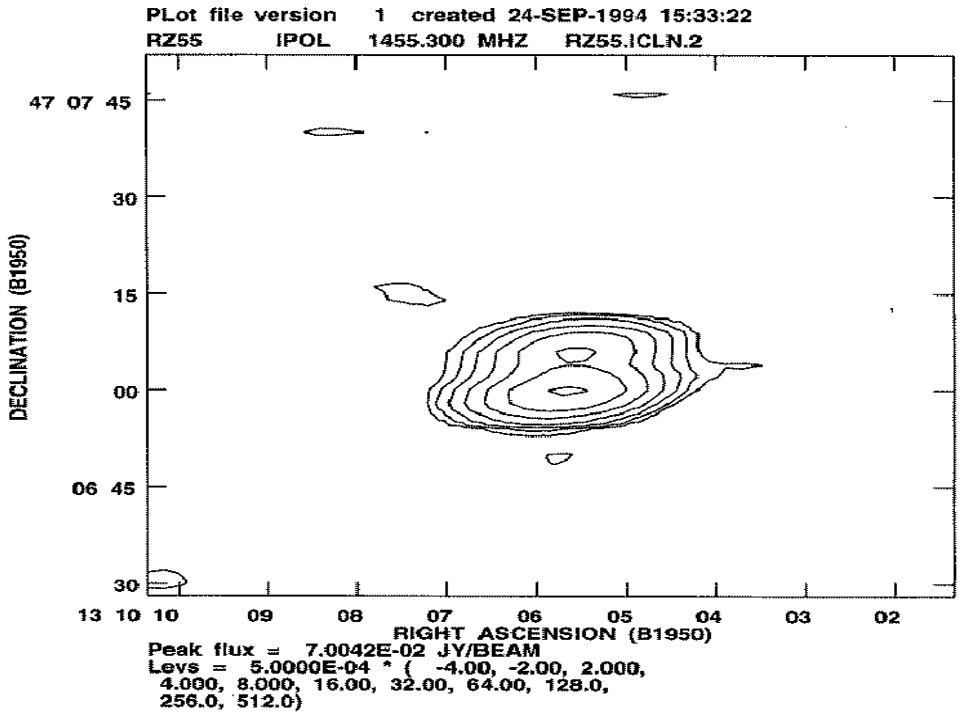,width=14cm,height=10cm}}
\caption{Radio source RZ55 (J~131217+465106).}
\end{figure*}

\begin{figure*}
\centerline{\psfig{figure=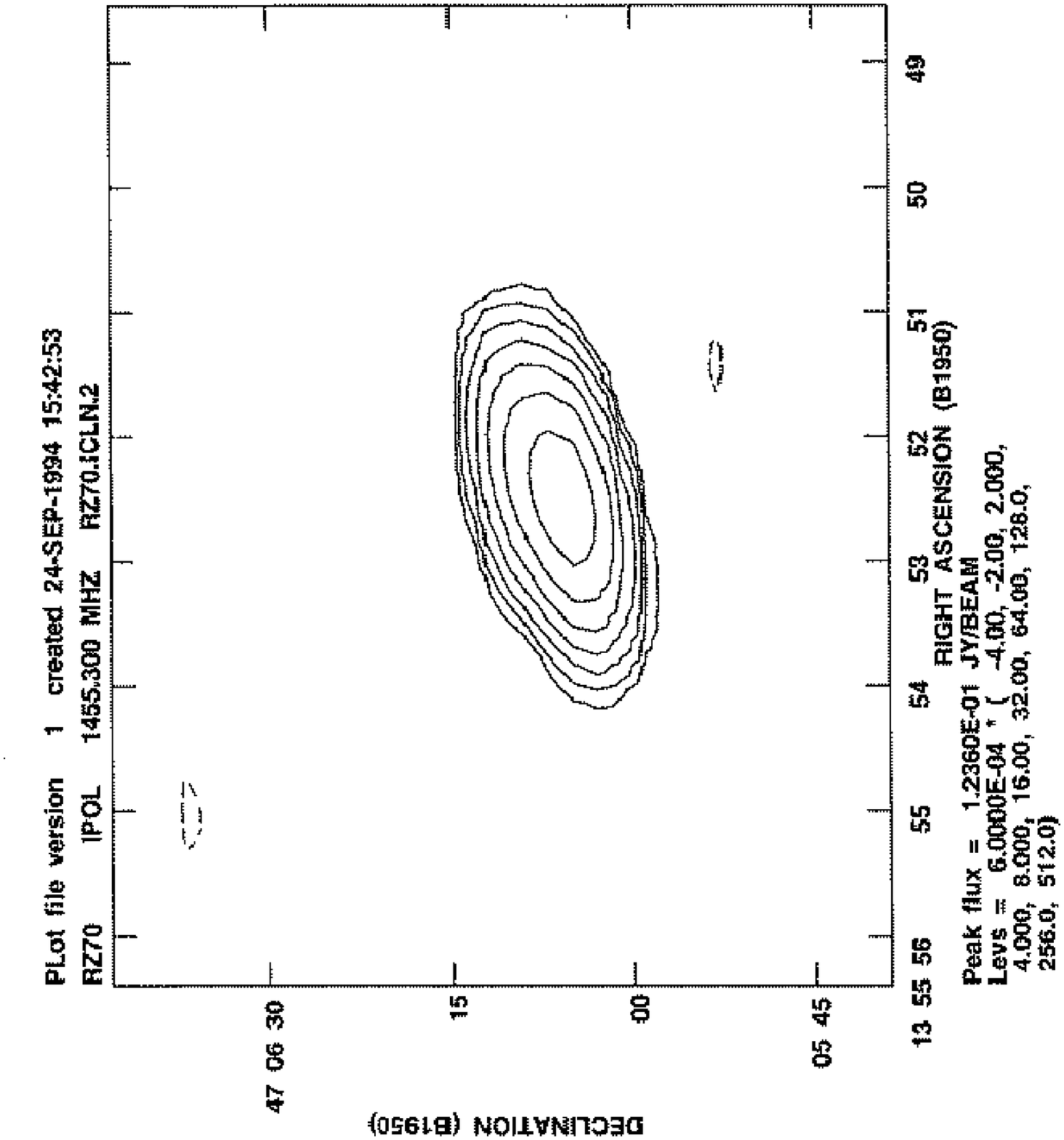,width=14cm,height=10cm,angle=-90}}
\caption{Radio source RZ70 (J~135751+465130).}
\end{figure*}

\vspace{0.1cm}
\section{Radio source spectra}

To prepare the radio spectra of the sources, we have used the catalogs of
the database CATS (astrophysical CATalogs Support system)
(Verkhodanov et al., 1997) and the data of the survey carried out at the
North sector of RATAN--600 in 1995
(Verkhodanov and Verkhodanova, 1998, 1999).
\clearpage
\begin{onecolumn}

\begin{center}
\topcaption{Multifrequency data for 5 RZ radio sources. Fitting curves
are derived with $y = log S$ and $x = log \nu$, where
$S$ is the flux density in Jy, $\nu$ is the frequency in MHz}
\tablefirsthead{\hline
		    \multicolumn{1}{|c}{$\alpha$}
		 &  \multicolumn{1}{c}{$\sigma_\alpha$}
		 &  \multicolumn{1}{c}{$\delta$}
		 &  \multicolumn{1}{c}{$\sigma_\delta$}
		 &  \multicolumn{1}{c}{$\nu$}
		 &  \multicolumn{1}{c}{S}
		 &  \multicolumn{1}{c}{$\sigma_S$}
		 &  \multicolumn{1}{c|}{Catalog}        \\
\hline
		    \multicolumn{1}{|c}{$~~^h~~^m~~^s$}
		 &  \multicolumn{1}{c}{$~~^s$}
		 &  \multicolumn{1}{c}{~~\degr~\arcmin~~\arcsec}
		 &  \multicolumn{1}{c}{~~\arcsec}
		 &  \multicolumn{1}{c}{MHz}
		 &  \multicolumn{1}{c}{Jy}
		 &  \multicolumn{1}{c}{Jy}
		 &  \multicolumn{1}{c|}{}            \\
\hline
		    \multicolumn{1}{|c}{1}
		 &  \multicolumn{1}{c}{2}
		 &  \multicolumn{1}{c}{3}
		 &  \multicolumn{1}{c}{4}
		 &  \multicolumn{1}{c}{5}
		 &  \multicolumn{1}{c}{6}
		 &  \multicolumn{1}{c}{7}
		 &  \multicolumn{1}{c|}{8}           \\
 \hline              }
\tablehead{ \hline
		    \multicolumn{1}{|c}{$~~^h~~^m~~^s$}
		 &  \multicolumn{1}{c}{$~~^s$}
		 &  \multicolumn{1}{c}{~~\degr~\arcmin~~\arcsec}
		 &  \multicolumn{1}{c}{~~\arcsec}
		 &  \multicolumn{1}{c}{MHz}
		 &  \multicolumn{1}{c}{Jy}
		 &  \multicolumn{1}{c}{Jy}
		 &  \multicolumn{1}{c|}{}            \\
		    \multicolumn{1}{|c}{1}
		 &  \multicolumn{1}{c}{2}
		 &  \multicolumn{1}{c}{3}
		 &  \multicolumn{1}{c}{4}
		 &  \multicolumn{1}{c}{5}
		 &  \multicolumn{1}{c}{6}
		 &  \multicolumn{1}{c}{7}
		 &  \multicolumn{1}{c|}{8}            \\
     \hline          }
\tabletail{\hline}
\begin{supertabular}{|lllrrlll|}
	     &       &             &      &      &       &         &       \\
\multicolumn{8}{|l|}
	{{\bf RZ 5}\hspace*{1cm}   $y = 1.478 - 0.736x$} \\
08 23 46.2   &   0.5 &+46 51 40    &    5 &  151 &   0.72&    0.04 & 6CVI  \\
08 23 45     &   1.8 &+46 51 56.5  & 24.7 &  232 &   0.28&    0.05 & MIYUN \\
08 23 47.0 3 &       &+46 51 52.7  &      &  325 &  0.465&  0.0055 & WENSS \\
08 23 46.779 & 0.121 &+46 51 50.48 & 0.62 &  365 &  0.445&   0.042 & TXS   \\
08 23 46.8   &       &+46 51 54    &      &  408 &   0.38&    0.05 & B3    \\
08 23 47.13  &  0.04 &+46 51 50.4  &  0.6 & 1400 & 0.1522&  0.0054 & NVSS  \\
08 23 46.84  &       &+46 51 53.05 &      & 1400 &  0.128&         & WB92  \\
08 23 46.18  &  0.02 &+46 52 00.3  &  0.2 & 1400 &0.03357& 0.00014 & FIRST \\
08 23 47.37  &  0.01 &+46 51 48.6  &  0.1 & 1400 &0.11003& 0.00014 & FIRST \\
08 23 50     &       &+46 51 18    &      & 1400 &   0.17&         & GB    \\
08 23 47.9   &   0.2 &+46 52 09    &   15 & 2308 &  0.123&   0.033 & RATAN \\
08 23 47.1   &   0.1 &+46 51 55    &    2 & 3750 &  0.055&   0.006 & RZ    \\
08 23 47.9   &   0.1 &+46 52 09    &   15 & 3950 &  0.052&   0.008 & RATAN \\
08 23 46.0   &   0.8 &+46 51 51    &    9 & 4850 &  0.075&   0.007 & GB6   \\
08 23 46.7   &   1.1 &+46 51 56    &   12 & 4850 &  0.069&   0.009 & 87GB  \\
08 23 47.9   &   0.2 &+46 52 09    &   15 & 7700 &   0.05&   0.010 & RATAN \\
08 23 47.9   &   0.2 &+46 52 09    &   15 &11111 &  0.035&   0.010 & RATAN \\
	     &       &             &      &      &       &         &       \\
\multicolumn{8}{|l|}
	 {{\bf RZ 9}\hspace*{1cm} $y = -1.663+0.001x+11.190exp(-x)$} \\
08 41 43.0  &   0.5 &+46 52 08  &    5 &  151 &    0.37 &   0.02 & 6CVI   \\
08 41 44.9  &   0.5 &+46 52 19  &    5 &  151 &    0.41 &   0.02 & 6CII   \\
08 41 41.15 &       &+46 52 33.9&      &  325 &   0.182 & 0.0036 & WENSS  \\
08 41 41    &   0.2 &+46 52 52  &  1.7 &  327 & 0.19531 &0.00834 & WSRTW  \\
08 41 40.3  &       &+46 52 52  &      &  408 &    0.12 &   0.05 & B3     \\
08 41 41.24 &  0.01 &+46 52 34.7&  0.1 & 1400 & 0.07445 &0.00014 & FIRST  \\
08 41 41.28 &  0.05 &+46 52 34.9&  0.6 & 1400 &  0.0737 & 0.0016 & NVSS   \\
08 41 41.8  &   0.1 &+46 52 50  &    2 & 3750 &   0.044 &  0.006 & RZ     \\
08 41 42.1  &   0.2 &+46 53 04  &   15 & 3950 &   0.043 &  0.008 & RATAN  \\
08 41 41.9  &   1.1 &+46 52 56  &   12 & 4850 &   0.035 &  0.005 & GB6    \\
08 41 42.4  &   1.5 &+46 53 43  &   17 & 4850 &   0.057 &  0.006 & 87GB   \\
08 41 42.1  &   0.2 &+46 53 04  &   15 & 7700 &   0.039 &  0.010 & RATAN  \\
	    &       &           &      &      &         &        &        \\
\multicolumn{8}{|l|}
	    {{\bf RZ 14}\hspace*{1cm} $y = 1.512 - 0.914x$}\\
08 48 19.6   &  0.5 &+46 52 34    &    5 &  151 &     0.3 &   0.02 & 6CVI  \\
08 48 22.5   &  0.5 &+46 51 54    &    5 &  151 &    0.37 &   0.02 & 6CII  \\
08 48 18.18  &      &+46 51 53.9  &      &  325 &    0.17 & 0.0036 & WENSS \\
08 48 18.6   &      &+46 51 42    &      &  408 &    0.12 &   0.05 & B3    \\
08 48 18.21  & 0.01 &+46 51 53.1  &  0.1 & 1400 & 0.05126 &0.00014 & FIRST \\
08 48 18.24  & 0.05 &+46 51 53.2  &  0.6 & 1400 &  0.0503 & 0.0008 & NVSS  \\
08 48 22.5   &  0.3 &+46 52 21    &    4 & 3750 &   0.016 &  0.006 & RZ    \\
08 48 17.6   &  0.2 &+46 52 01    &   15 & 3950 &    0.03 &  0.008 & RATAN \\
	     &      &             &      &      &         &        &       \\
\multicolumn{8}{|l|}
	    {{\bf RZ 55}\hspace*{1cm} $y = 1.470 - 0.775x$} \\
13 12 18.8   &  0.5 &+46 51 22    &    5 &   151 &     0.6&    0.03 & 6CII  \\
13 12 18     &  1.4 &+46 51 04.7  & 18.8 &   232 &    0.51&    0.05 & MIYUN \\
13 12 17.22  &      &+46 51 09.5  &      &   325 &    0.36&  0.0036 & WENSS \\
13 12 17.297 &0.153 &+46 51 07.77 & 0.67 &   365 &    0.28&   0.018 & TXS   \\
13 12 16.9   &      &+46 51 08    &      &   408 &    0.29&    0.05 & B3    \\
13 12 17.40  & 0.01 &+46 51 07.4  &  0.1 &  1400 & 0.11988& 0.00013 & FIRST \\
13 12 17.43  & 0.05 &+46 51 07.9  &  0.6 &  1400 &  0.1146&  0.0005 & NVSS  \\
13 12 22     &      &+46 51 42    &      &  1400 &    0.09&         & GB    \\
13 12 20.1   &  0.1 &+46 51 06    &    2 &  3750 &   0.035&   0.006 & RZ    \\
13 12 18.9   &  0.2 &+46 52 20    &   15 &  3950 &    0.04&   0.008 & RATAN \\
13 12 17.0   &    1 &+46 51 18    &   11 &  4850 &    0.05&   0.005 & GB6   \\
13 12 17.3   &  1.6 &+46 51 19    &   22 &  4850 &   0.053&   0.007 & 87GB  \\
13 12 18.9   &  0.2 &+46 52 20    &   15 & 11111 &   0.046&   0.010 & RATAN \\
	     &      &             &      &       &         &        &       \\
\multicolumn{8}{|l|}
	 {{\bf RZ 70}\hspace*{1cm} $y = 2.113 - 0.936x$} \\
13 57 51.1  &   0.5 &+46 51 24   &    5 &   151 &    1.23 &   0.06 & 6CII  \\
13 57 50.6  &   0.8 &+46 52 02   & 10.8 &   232 &    0.67 &   0.05 & MIYUN \\
13 57 53.1  &   0.9 &+46 51 39.6 &   12 &   232 &     0.6 &   0.05 & MIYUN \\
13 57 51.04 &       &+46 51 32.8 &      &   325 &   0.607 & 0.0031 & WENSS \\
13 57 51.164& 0.089 &+46 51 31.86& 0.36 &   365 &   0.471 &  0.016 & TXS   \\
13 57 51.4  &       &+46 51 36   &      &   408 &    0.46 &   0.05 & B3    \\
13 57 51.13 &   0.1 &+46 51 31.45& 0.01 &  1400 & 0.16628 &0.00014 & FIRST \\
13 57 51.2  &       &+46 51 30.3 &      &  1400 &    0.14 &        & UCC R \\
13 57 51    &       &+46 52 20   &      &  1400 &    0.15 &        & GB    \\
13 57 49.34 &       &+46 51 51.45&      &  1400 &   0.161 &        & WB92  \\
13 57 51.11 &  0.04 &+46 51 31.87&  0.6 &  1400 &  0.1674 & 0.0004 & NVSS  \\
13 57 54.0  &   0.2 &+46 52 34   &   15 &  2308 &   0.075 &  0.033 & RATAN \\
13 57 54.6  &   0.1 &+46 51 29   &    2 &  3750 &   0.024 &  0.006 & RZ    \\
13 57 54.0  &   0.2 &+46 52 34   &   15 &  3950 &   0.068 &  0.008 & RATAN \\
13 57 51.3  &   1.1 &+46 51 37   &   13 &  4850 &   0.039 &  0.005 & GB6   \\
13 57 49.9  &   1.9 &+46 51 47   &   25 &  4850 &   0.035 &  0.006 & 87GB  \\
13 57 54.0  &   0.2 &+46 52 34   &   15 &  7700 &   0.054 &  0.010 & RATAN \\
13 57 54.0  &   0.2 &+46 52 34   &   15 & 11111 &   0.037 &  0.010 & RATAN \\
\end{supertabular}
\end{center}

\end{onecolumn}

Among the basic catalogs stored in the CATS database several
largest ones (NVSS, FIRST, WENSS, Texas) cover entirely  the area
of the Zenith survey.

Using these catalogs we have detected
the five objects to have counterparts in the NVSS
(Condon et al., 1998), which has a sensitivity
as high as 2.5\,mJy, and a resolution of 45\arcsec ~at 1400~MHz.

\begin{figure*}
\centerline{\psfig{figure=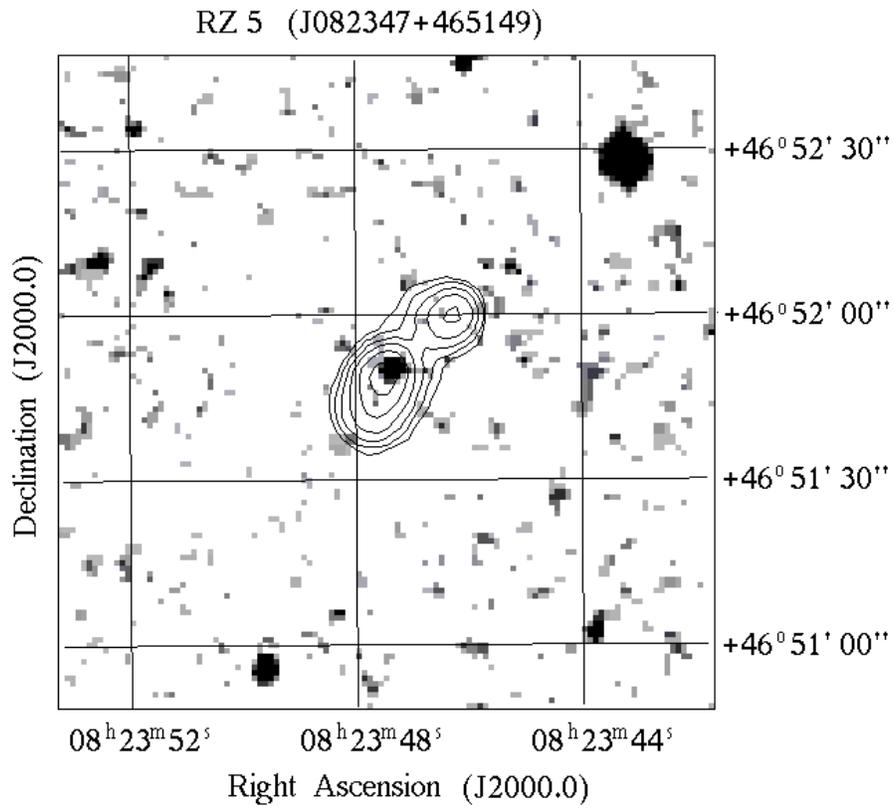,width=12cm,bbllx=1pt,bblly=1pt,bburx=660pt,bbury=615pt,clip=}}
\caption{The FIRST map of the radio source RZ5 (J~082347+465150)
overlaid on the 2\arcmin$\times$2\arcmin DSS image.}
\end{figure*}

\begin{figure*}
\centerline{\psfig{figure=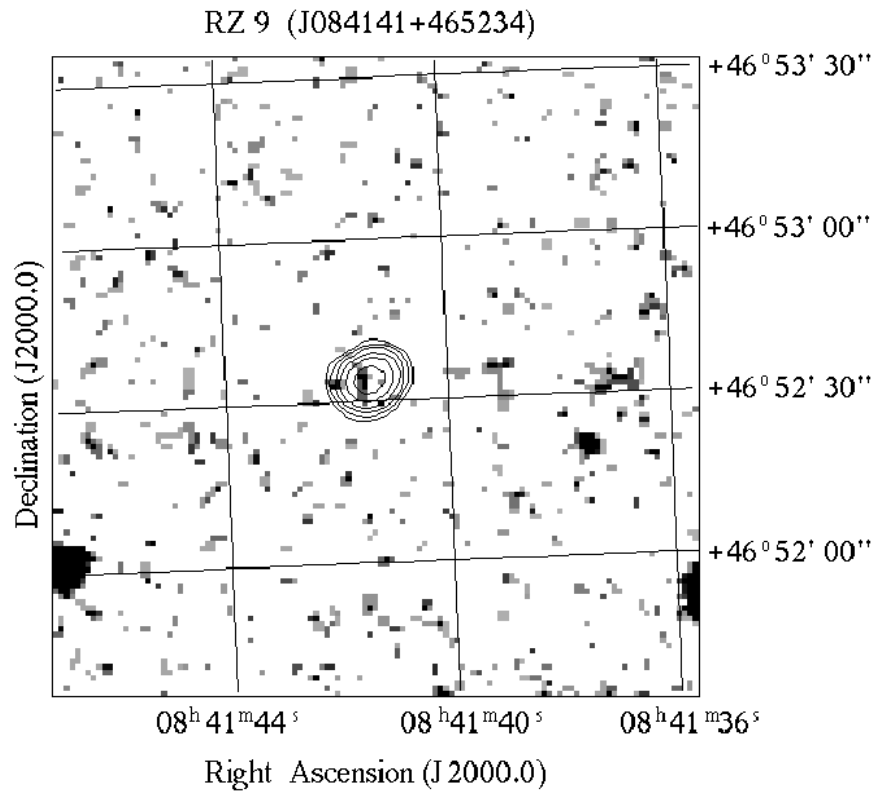,width=12cm}}
\caption{The FIRST image of the radio source RZ9 (J~084141+465234)
overlaid on the 2\arcmin$\times$2\arcmin DSS image.}
\end{figure*}

\begin{figure*}
\centerline{\psfig{figure=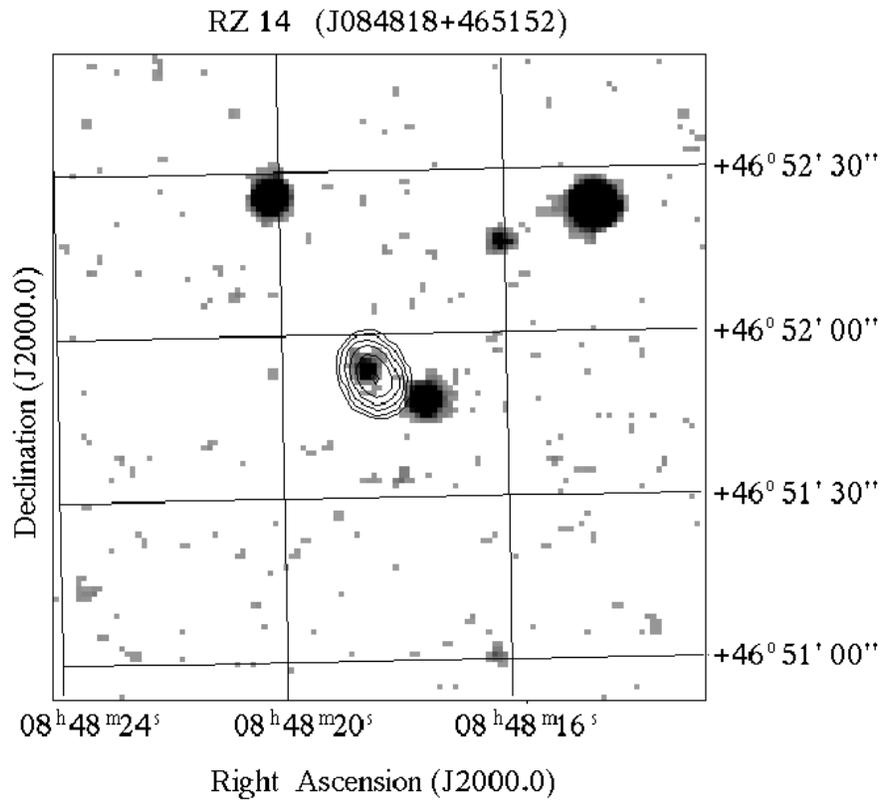,width=12cm}}
\caption{The FIRST image of the radio source RZ14 (J~084818+465153)
overlaid on the 2\arcmin$\times$2\arcmin DSS image.}
\end{figure*}

\begin{figure*}
\centerline{\psfig{figure=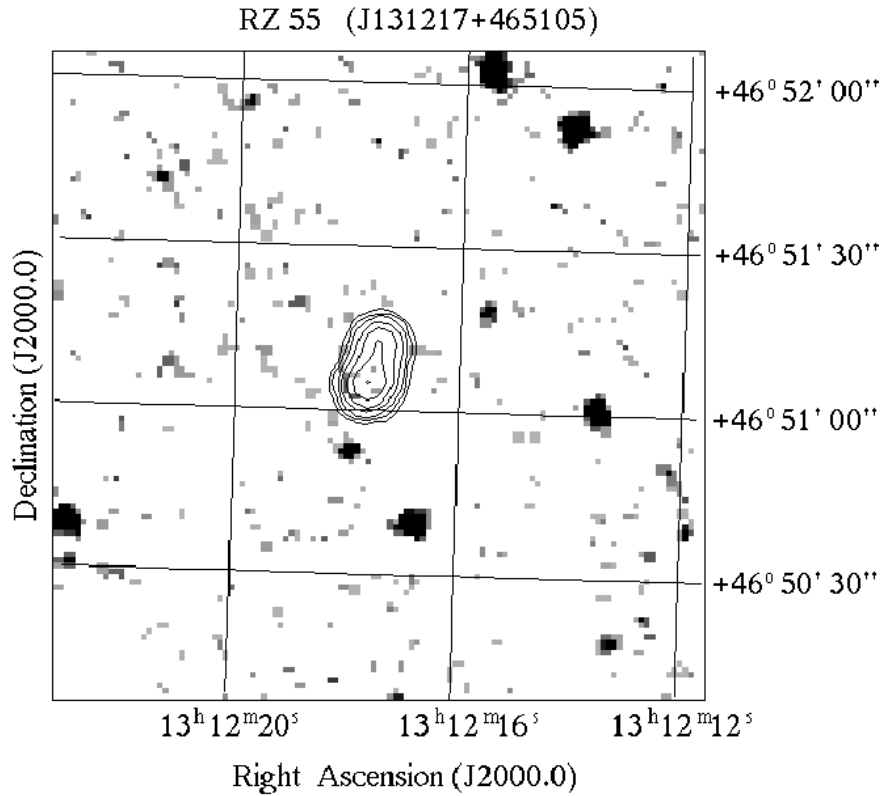,width=12cm}}
\caption{The FIRST image of the radio source RZ55 (J~131217+465106)
overlaid on the 2\arcmin$\times$2\arcmin DSS image.}
\end{figure*}

\begin{figure*}
\centerline{\psfig{figure=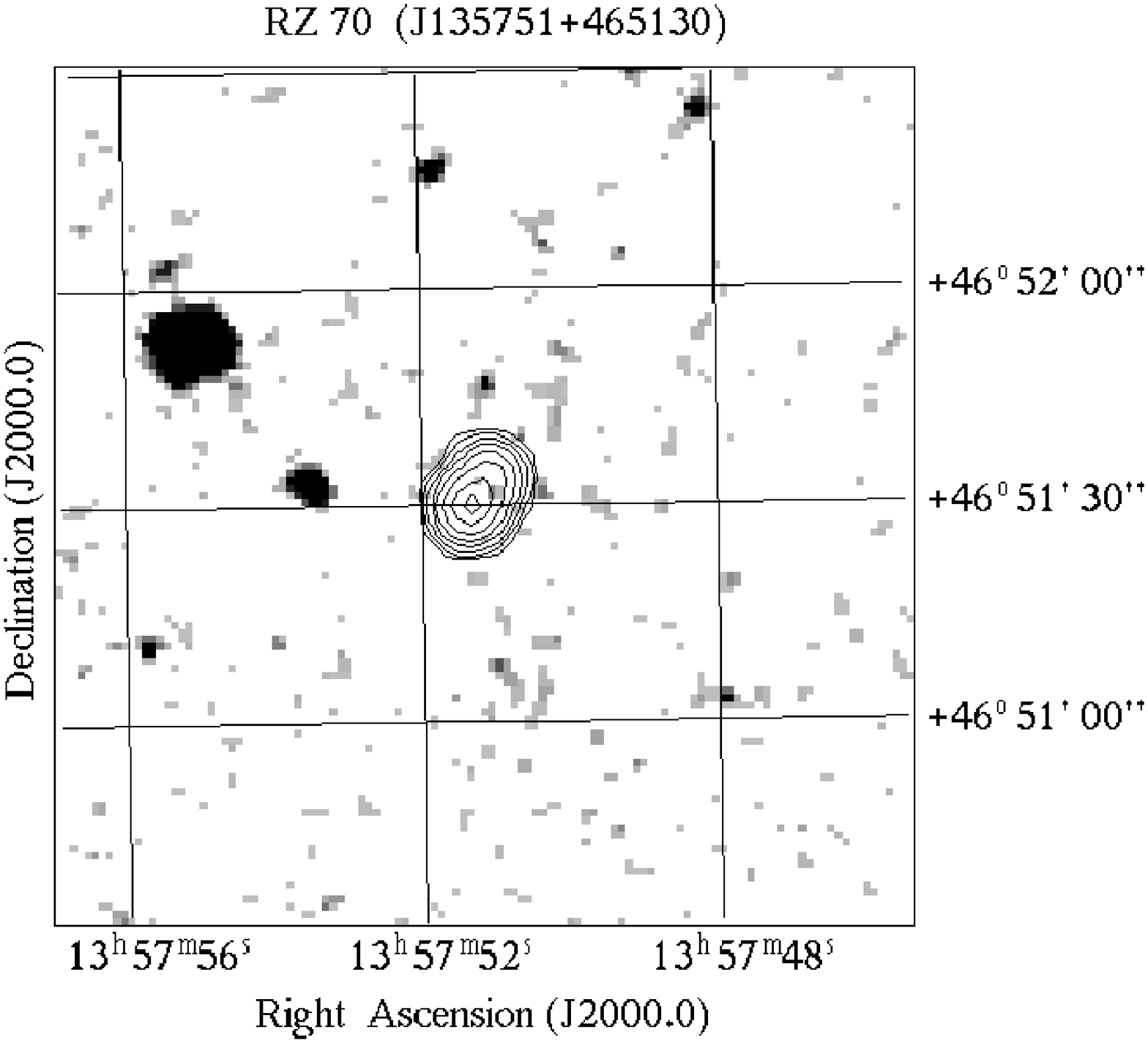,width=12cm}}
\caption{The FIRST image of the radio source RZ70 (J~135751+465130)
overlaid on the 2\arcmin$\times$2\arcmin DSS image.}
\end{figure*}

In the FIRST survey
(White et al., 1997) which is carried out at VLA
in the B--configuration at 1400\,MHz with a resolution of
5\farcs4 and a sensitivity limit of 1 mJy at  5$\sigma$ level,
we have detected all five {\it RZ} sources.
The coordinates of the objects coincide with those of 1994 VLA observations
within 0$\farcs$05. All the basic components of the complex
sources (except the tail of the object {\it RZ9} (Fig.2))
are confirmed by the FIRST data.
Images of the FIRST sources are shown in Figs. 6--10.

The WENSS survey
(Rengelink et al., 1997)
carried out at 325 MHz with a sensitivity of about 18\,mJy at
5$\sigma$ level and a resolution of
$54\arcsec\times{54}\arcsec \cdot {\rm cosec} \delta$
has counterparts for all five sources.
The Texas survey carried out at 365 MHz on the Texas radio interferometer
(Douglas et al., 1996) with a sensitivity as high as 150 mJy,
but complete to 250 mJy, contains counterparts of
only three sources:
{\it RZ5}, {\it RZ55} and {\it RZ70}.

Table 2 contains a list of identifications with the catalogs of the CATS
database (Verkhodanov et al., 1997). The data of the $RZ$ catalog have been
taken from Mingaliev et al. (1991) and the ones marked as RATAN
are from the paper by Verkhodanov and Verkhodanova (1999).

The radio spectra have been fitted with the package {\it SPG}
under OS Linux (Verkhodanov, 1997).
As the main fitting function of the radio spectra the authors have used
the curve $y = A + Bx$,
where $x = log~\nu$, $y = log~S$, $\nu$ is the frequency in MHz,
$S$ is a flux density in Jy.

The fitting procedure took into account weights of each point
in proportion to the value
$1/\epsilon^2$, where $\epsilon$ is the ratio of the  flux density error
$\sigma_S$ and the flux density $S$. If a point of the spectrum
is appreciably away from
the spectrum line, its weight is decreased  ten
times.

Automatic selection using the least square criterion
from the fitted curve taken from the set\\
     $y = A + Bx$,\\
     $y = A + Bx + Cx^2$,\\
     $y = A + Bx + C \times {\rm exp} (x)$,\\
     $y = A + Bx + C \times {\rm exp} (-x)$,\\
\noindent shows the linear fitting for four sources, and for $RZ9$
with the curve $y = -1.663+0.001x+11.190{\rm exp}(-x)$.
In the case of linear approximation, the best fit for $RZ9$
radio spectrum is done with the line $y = 1.024\,-\,0.678x$.
When we fitted the $RZ9$ spectrum, we took into account
that a corresponding point of the 87GB survey is likely to be wrong and
the RATAN point
at 2.7~cm may be too high because of the noise.

The steep linear spectrum due to synchrotron radiation
is typical of radio galaxies and some quasars.
If the object $RZ$9 has a concave spectrum, it could be explained
in the frames of a model of superposition of two spectra:
a flat spectrum of a core, and a steep spectrum of component(s).

The spectra of the radio sources are shown in Fig.11.

\section {The search for optical candidates}

Electronic versions of the Palomar Observatory Sky Survey
have been used to search for candidates for optical identification.
We used the APM catalog
(see, e.g. Irwin, 1998), to be exact, the modified
client program of T.\,McGlynn {\it apmcat}, for the stream identification
of the sources via Internet and estimation of magnitudes in the R and B filters,
and the DSS2 (Digitized Sky Survey), accessible via the Web-page
of the Space Telescope Science Institute (http://stdatu.stsci.edu/dss/)
for the identification of the Zenith survey sources.

DSS frames of 2\arcmin$\times$2\arcmin~ in size with
the overlaid FIRST
maps are shown in Figs. 6--10 for all five sources.
The candidates for optical identification of the first three sources
are situated inside the central isophote.
There are no optical candidates brighter than 21$^m$ in E--band for $RZ55$ and
$RZ70$ on DSS.

\begin{table*}
\caption{Results of optical identifcation}
\begin{center}
\begin{tabular}{|lcclrrrr|}
\hline
Name &$\alpha+\delta$ (radio)& $\alpha+\delta$ (APM)&\hs E & Cl & O  & Cl & LR  \\
\hline
RZ5C& 082347.13+465150.4&082347.14+465152.0&\hs19.53  &-1&$>$21.5\hs&0&93.39\\
RZ5N& 082346.18+465200.3&   -"-       &\hs-"- &-"-&-"-&-"-& 0.00\\
RZ5S& 082347.37+465148.6&   -"-       &\hs-"- &-"-&-"-&-"-& 0.07\\
RZ9 & 084141.23+465234.5&084141.46+465236.3&$>$21$^1$ & 0  &21.13& 1 &4.43 \\
RZ14& 084818.21+465153.1&084818.33+465154.8&\hs19.03  &$-1$  &19.26&$-1$ & 35.37\\
RZ55& 131217.54+465106.0&                  &$>$21$^1$ &    &$>$21.5\hs& & \\
RZ70& 135751.31+465130.5&                  &$>$24$^2$ &    &$>$21.5\hs& & \\
\hline
\end{tabular}
\end{center}
\begin{flushleft}
$^1$  DSS2 magnitude values\\
$^2$  6 m telescope data
\end{flushleft}
\end{table*}

\begin{figure*}
\centerline{
\psfig{figure=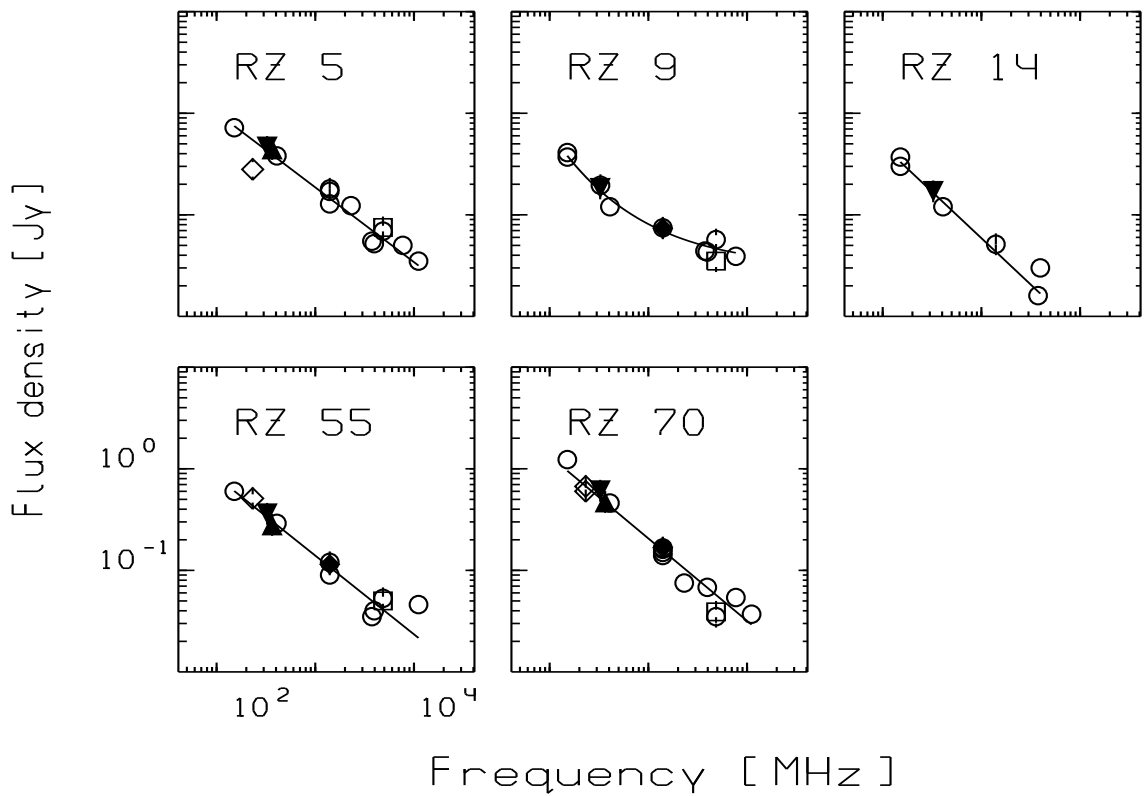,width=16cm,angle=-90,bbllx=40pt,bblly=80pt,bburx=450pt,bbury=650pt,clip=}}
\caption{The spectra of the radio sources.}
\end{figure*}

\begin{figure*}
\centerline{\psfig{figure=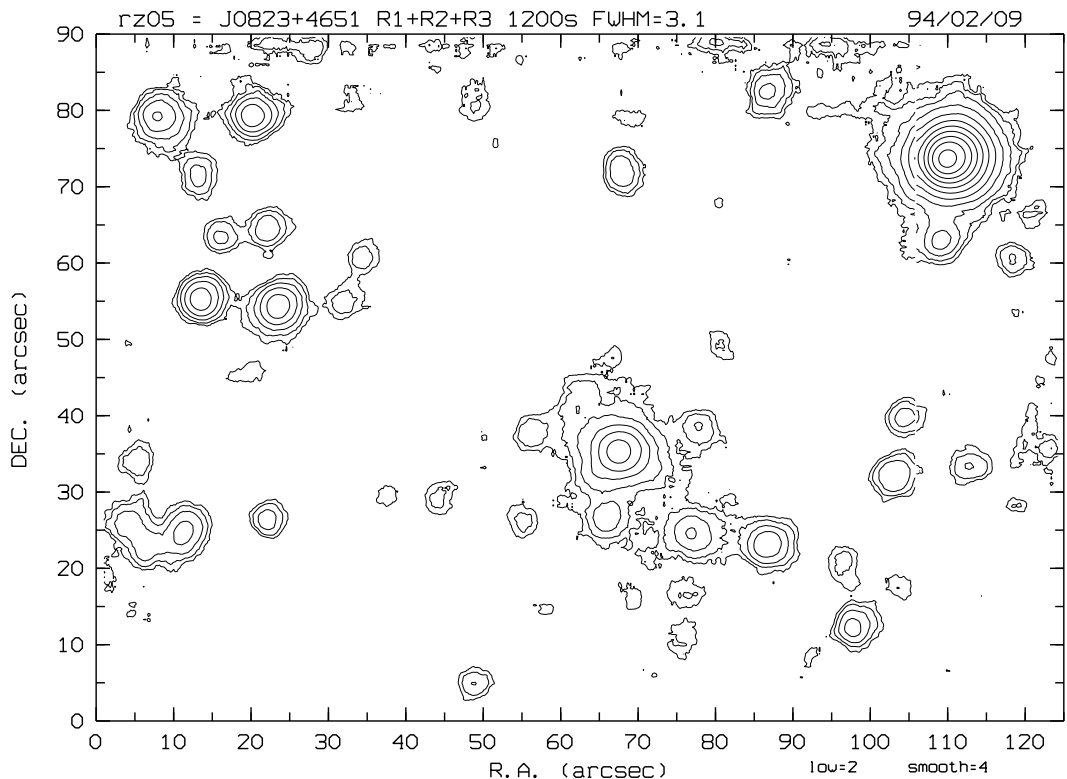,width=16cm,angle=-90}}
\caption{The map of the 6m telescope CCD image of radio source RZ5
(J~082347+465150).}
\end{figure*}
Two radio sources $RZ5$ and $RZ70$ were observed with the 6 m telescope
in February, 1994.
Three 400$^s$ CCD frames were acquired for $RZ5$ and six 400$^s$ ones
for $RZ70$ with a seeing of 3\farcs1.
Isophotes of the $RZ5$ CCD image are shown in Fig. 12.
There is an empty field in the zone of $RZ70$ in the R band.

The results of identification with the APM catalog and DSS and the upper limits
estimated with the  6 m telescope data are given in Table 3.

\section{Discussion}

Table 3 contains the results of optical identification of the objects with
the APM data. In the columns are listed
the object name (for {\it RZ5} the northern and southern components,
and a gravity center by NVSS), APM coordinates at 2000.0, the magnitude and
object class on the  E plate, the  magnitude and object class
on the O plate (the class is described in terms of APM, with $-1$ (stellar like),
0 (noise like)
and 1 (extended)), likelihood ratio, calculated by the following formula
(de Ruiter et al., 1977):
$$
  LR(r) = (1/2\lambda) \exp [0.5r^2(2\lambda-1)],
$$
where $\lambda = \pi\sigma_{RA}\sigma_{Dec}\rho$,
$\rho$ is the density of background objects
equal to 5.16$\times$10$^{-4}$ sec$^{-2}$
(Cohen et al., 1977),
$r = [(\Delta{RA}/\sigma_{RA})^2 + (\Delta{Dec}/\sigma_{Dec})^2]^{0.5}$,
$\Delta{RA}, \Delta{Dec}$ are differences of radio and optical positions,
$\sigma{RA}$ and
$\sigma{Dec}$ are root-mean-square errors of the radio and
optical coordinates, respectively.
The identification is considered  reliable if LR$>$2.

Analyzing the VLA maps (see Table 1) one can find that
all five objects
are not point-like.
Two objects ({\it RZ5} and {\it RZ55}) have distinguished
radio components. Identification of the objects with DSS and APM shows
only 3 radio sources identified with a high probability with
the optical objects visible on DSS. Despite the fact that {\it RZ5} is
identified
with a red star--like object of 19{\fm}5,
BTA observations of 1994 have shown that it is rather an elliptical
galaxy. An averaged 6 m CCD image (Fig. 12) of {\it RZ5} shows that
merging of galaxies is probably observed in this object.
Actually, it is not  merging of equal galaxies, but we have a case
of ``cannibalism'', i.e. a giant galaxy ``eats up'' surrounding galaxies
of lower mass.
Radio emission of this merged group of at least 8 galaxies, with the largest
one at the center, may be produced by  active processes working during
the merging, which may have an explosive character
(Kontorovich et al., 1992).

The source {\it RZ9}, which is slightly extended at 1400 MHz,
is identified with an extended one in optics (O-plate) and, may be,
with a galaxy or a QSO, i.e. a radio core.
If the 2\farcs5-wide tail on the left side of $RZ5$ (Fig. 2) is not a false
structure, then it may be a jet flying out from the core.

The source {\it RZ14}, having an extended but unresolved into components
structure
and a steep radio spectrum, coincides with a star-like object
with a very blue O-E color index, which is most likely a QSO
(blue star--like object, BSO). Taking into account that it has
an extended structure (probably two close radio components) we can suppose
it to be a candidate for a gravitationally lensed object (Fletcher, 1998).

The sources {\it RZ55} and {\it RZ70}, showing an extended structure
(although {\it RZ70} is not resolved into components) and having steep
spectra, are not visible on the Palomar Observatory Sky Survey plates.
Therefore, they may be selected as candidates for  distant radio
galaxies and  objects for further investigation.

\section{Conclusions}

VLA maps with a resolution of 2\farcs5$\times$2\arcsec~ for
three RATAN--600 Zenith radio sources,
{\it RZ5}, {\it RZ9} and {\it RZ14}, and
with a resolution of 6\farcs5$\times$2\farcs3~
for  two sources, {\it RZ55}, {\it RZ70}, have been analyzed.

All five objects are extended and two of them ($RZ5$ and $RZ55$)
are clearly resolved. All the sources have linear steep spectra
($\alpha < -0.65$), and one of them ($RZ14$) has an ultrasteep spectrum.

Confident optical counterparts have been found
in the Palomar Digitized Sky Survey for 3 radio sources.

The radio source $RZ5$, being a merging group of one large and
several small galaxies (as 6\,m telescope observations show), may have appeared
in this process.

The radio source $RZ14$ is probably a BSO and could be a candidate
to a gravitationally lensed object.

Two radio sources, {\it RZ55} and {\it RZ70}, which have
no optical counterparts up to 21\fm5 ($RZ70$
up to 24$^m$), may be distant radio galaxies ($z >$ 0.5).

\begin{acknowledgements}
The authors are grateful to S.A.Pustilnik for useful comments on this paper.
This work has been supported partly by
the Federal program ``Astronomy'' (grants 1.2.1.2 and
1.2.2.4), Federal program ``Integration'' (grant No 578),
and the Russian Foundation of Basic Research (grant No 99-07-90334).
\end{acknowledgements}


\begin{thebibliography}{}
\bibitem{}
Cohen A.M., Porcas R.W., Browne I.W., Daintree E.J., Walsh D., 1977.
    Mem. of Roy. Astr. Soc.,  {\bf 84}, 1

\bibitem{}
Condon J.J., Dickey J.M., Salpeter E.E., 1990, \aj, {\bf 99}, 1071

\bibitem{}
Condon J.J., Cotton W.D., Greisen E.W., Yin Q.F., Perley R.A.,
    Broderick J.J., 1998, \aj, {\bf 115}, 1693

\bibitem{}
Douglas J. N., Bash F. N., Bozyan F. A., Torrence G.W., Wolfe C., 1996,
    \aj, {\bf 111}, 1945

\bibitem{}
Fletcher A., 1998, Ph.D. thesis, Massachusetts Inst. of Tech. Boston, USA

\bibitem{}
Irwin Mike, 1998, \verb^http://www.ast.cam.ac.uk/~apmcat/^

\bibitem{}
Kontorovich~ V.M., ~Kats~ A.V.,~ Krivitskij D.S.,~ 1992, \newpage \noindent Pisma v Zhurnal
    Eksperimentalnoj i Teoreticheskoj Fiziki.
    {\bf 55}, No 1, 3
\bibitem{}
Mingaliev M.G., Verkhodanov O.V., Khabrakhmanov A.R, Temirova A.V.,
    Gol'nev V.Ya., 1991, Soobshch. Spets. Astrofiz. Obs., {\bf 68}, 47

\bibitem{}
Parijskij Yu. N., Goss W.M., Kopylov A.I., Soboleva N.S.,
    Temirova A.V.,Verkhodanov O.V., Zhelenkova O.P., Naugolnaya M.N.,
    1996, \bsao, {\bf 40}, 5

\bibitem{}
Rengelink R.B., Tang Y., de Bruyn A.G., Miley G.K.,
    Bremer M.N., Rottgering H.J.A., Bremer M.A.R., 1997,
    \aas, {\bf 124}, 259
    http://www.strw.leidenuniv.nl/wenss/

\bibitem{}
de Ruiter H.R., Willis A.G., Arp H.C., 1997, \aas,
    {\bf 28}, 211

\bibitem{}
Verkhodanov O.V., 1994, \azh, {\bf 71}. 352

\bibitem{}
Verkhodanov O.V., Trushkin S.A., Andernach H., Chernenkov V.N., 1997,
    Astronomical Data Analysis Software and Systems VI.
    ASP Conference Series, eds.: G.Hunt and H.E.Payne.
    {\bf 125}, 322
    http://cats.sao.ru

\bibitem{}
Verkhodanov O.V., 1997,
    In: Problems of modern radio astronomy, XXVII Radio Astron. Conf.
    St.-Petersburg. Inst. of Applied Astronomy, {\bf 1}, 322

\bibitem{}
Verkhodanov O.V., Verkhodanova N.V., 1997,
    In "Problems of modern radio astronomy". XXVII Radio astron. conf.
    St.-Petersburg. Inst. of Applied Astronomy, {\bf 1}, 195

\bibitem{}
Verkhodanov O.V., Verkhodanova N.V., 1998,
    In: Current problems of extragalctic astronomy,
    Proc. XV Conf., May 25--29, Pushchino,
    Pushchino Sci.Center, 18

\bibitem{}
Verkhodanov O.V., Verkhodanova N.V., 1999, \azh, {\bf 76}, No. 7 (in press)

\bibitem{}
White R.L., Becker R.H., Helfand D.J., Gregg M.D., 1997, \apj,
    {\bf 475}, 479

\end{thebibliography}
\end{document}